\documentclass[%
 reprint,
superscriptaddress,
 amsmath,amssymb,
 aps,
 pre,
]{revtex4-2}
\usepackage{xcolor}
\newcommand{\cblue }{\color{black}}

\usepackage{graphicx}
\usepackage{dcolumn}
\usepackage{bm}

\newcommand{\lla}{\left\langle}
\newcommand{\rra}{\right\rangle}

\begin{document}

\title{Activity enhanced shear-thinning of flexible linear polar polymers}
\author{Arindam Panda}%
\email{arindam19@iiserb.ac.in}
\affiliation{Department of Physics, Indian Institute Of Science Education and Research, \\ Bhopal 462 066, Madhya Pradesh, India}
\author{Roland G. Winkler}%
\email{r.winkler@fz-juelich.de}
\affiliation{Theoretical Physics of Living Matter, Institute for Advanced Simulation, Forschungszentrum J{\"u}lich, 52425 J{\"u}lich, Germany}
\author{Sunil P Singh}
 \email{spsingh@iiserb.ac.in}
\affiliation{Department of Physics, Indian Institute Of Science Education and Research, \\ Bhopal 462 066, Madhya Pradesh, India}



\date{\today}

\begin{abstract}
The rheological properties of tangentially propelled flexible polymers under linear shear flow are studied by computer simulations and are compared with analytical calculations. We find a significant impact of the coupled nonequilibrium active and shear forces on the polymer characteristics. The polar activity enhances shear-induced stretching along the flow direction, shrinkage in the transverse direction, and implies a strongly amplified shear-thinning behavior. The characteristic shear rate for the onset of these effects is determined by the activity. In the asymptotic limit of large activities, the shear-induced features become independent of activity, and for asymptotically large shear rates, shear dominates over activity with passive polymer behavior.          

\end{abstract}

\maketitle
\section{Introduction}

Active suspensions exhibit remarkable mechanical and flow responses, which differ significantly from those of their passive counterparts. Microswimmers, such as bacteria like {\it E. coli} and {\it Bacillus subtilis}~\cite{lauga2009hydrodynamics,elgeti2015physics}, exhibit a reduction in viscosity under shear flow~\cite{hatwalne2004rheology,sokolov2009reduction,gachelin2013non,lopez2015turning}, superfluidic like behavior~\cite{lopez2015turning,saintillan2018rheology}, alignment against flow and upstream~swimming~\cite{kaya2012direct,mathijssen2019upstream}.
The influence of flow on systems composed of extended filaments or polymer-like active structures can be even more complex because of their possible conformational changes. Biological cells comprise various ``active'' filaments and polymers, where activity is essential for a cell's function. Nonthermal fluctuations caused by DNA and RNA polymerases~\cite{belitsky2019polymerase} contribute to chromatin motion~\cite{weber2012chromosom,zidovska2013chromosom} and are involved in spatial segregation of the eukaryotic genome~\cite{lieberman2009chromosom,smrek2017chromosom,saintillan2018chromosom}.
Actin filaments and microtubules propelled by molecular motors determine the viscoelastic properties of the cytoskeleton~\cite{liverpool2001viscoelasticity} as well as that of the whole cell~\cite{pegoraro2017mechanical,chaudhuri2020viscoelastic,corominas2021viscoelastic}. Studies on flexible tubifex worms~\cite{deblais2020phase,deblais2020rheology} and semiflexible nematodes~\cite{backholm2013viscoelastic,malvar2019nematode} reveal an intriguing shear-thinning behavior and a significantly faster viscosity drop with increasing shear rate than passive polymeric fluids~\cite{xu2022nature}.

These examples highlight various features of active entities, especially polar filaments, and polymers, which are preferentially driving along their contour. This article focuses on the properties of flexible self-avoiding active polar linear polymers (APLPs) under linear shear flow in dilute solution. We aim to shed light on the simultaneous impact of the nonequilibrium forces, such as activity and shear, on a polymer's conformational and rheological properties.

The conformational, dynamical, and rheological properties of passive polymers under stationary shear flow have been intensively studied experimentally, theoretically, and by computer simulations~\cite{aust1999structure,doyle1997dynamic,smith1999single,larson2005rheology,schroeder2005dynamics,winkler2006semiflexible,hsu2010polymer,huang2010semidilute,singh2013dynamical}. 
Fluorescence light scattering experiments on DNA and actin filaments reveal a periodic stretching and recoiling dynamics -- denoted as tumbling motion --, which is accompanied by significant conformational fluctuations, rather than maintaining a stationary stretched state~\cite{schroeder2005dynamics,hur2001dynamics,larson2005rheology}. The average polymer extension increases with increasing shear rate along the flow direction, and its transverse extension shrinks~\cite{teixeira2005shear,saha2012tumbling,winkler2010conformational,huang2010semidilute}, which changes the tumbling frequency. This is accompanied by shear thinning, i.e., the shear viscosity decreases with increasing shear rate \cite{bird1987dynamics}.    
Molecular dynamics simulations provide detailed insight into nonequilibrium polymer properties, and confirm the experimental observations~\cite{huang2010semidilute,huang2012non,larson2005rheology,singh2013dynamical}. In particular, the tumbling frequency increases by a power law with an exponent of 2/3~\cite{huang2012non,singh2020flow,saha2012tumbling} for high shear rates. Similarly, the shear viscosity decreases in a power-law manner with an exponent of approximately $1/2$~\cite{schroeder2005shear,huang2012non,larson2005rheology,singh2013dynamical}.   

Flows are omnipresent in active matter systems, either induced by the physical environment~\cite{stocker2019flow} or by external flows. The latter will play a major role in the processing of advanced active matter systems, where the rheological features of the individual entities and their collective behavior will be of paramount importance. 

\begin{figure}[t]
    \centering
    \includegraphics[width=\columnwidth]{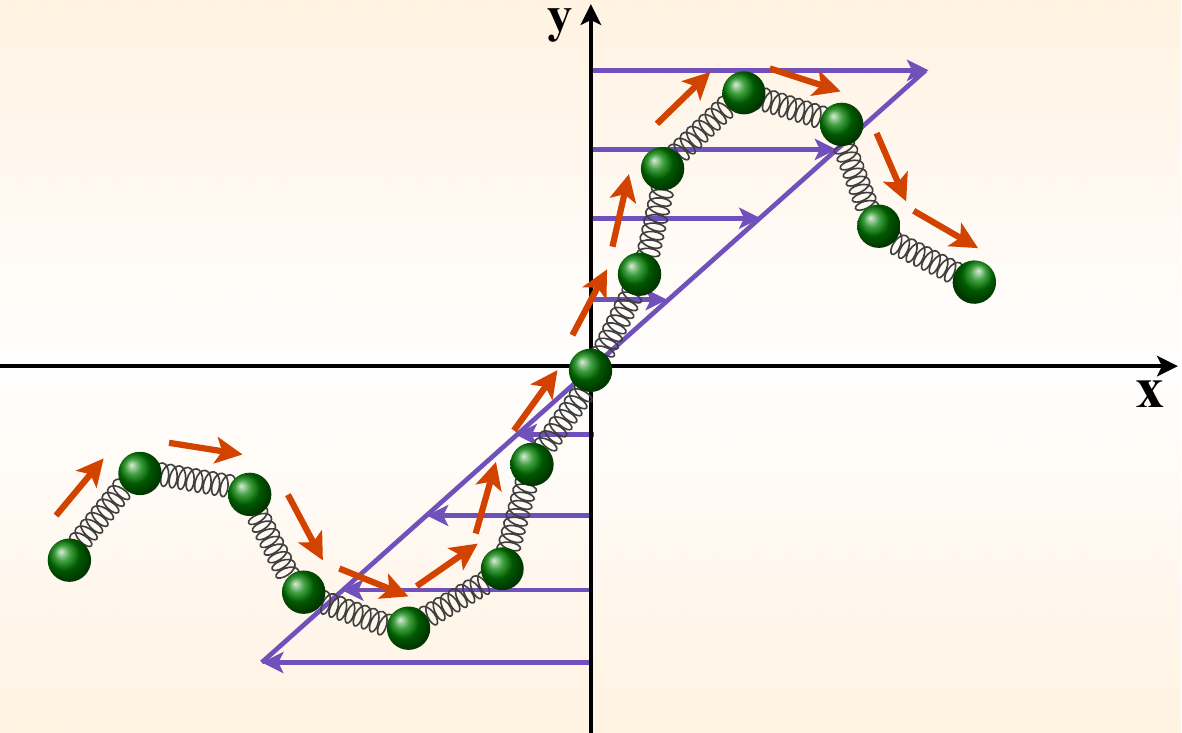}
    \caption{Schematic of a tangentially driven active polar linear polymer under linear shear flow. The red arrows indicate the active forces along the bond vectors. The purple arrows and the background color illustrate the linear shear flow.}
    \label{fig:zeroshear}
\end{figure}

Applying a coarse-grained, generic approach, which mimics the self-propulsion of microtubules, actin filaments, and worms~\cite{isele2015self,patra2022collective,bianco2018globulelike,anand2018structure,winkler2020physics,locatelli2021activity,philipps2022tangentially}, our simulations reveal a competition between the polar activity and the shear flow, with a significant impact on the polymer's conformations, dynamics, and rheological properties. Activity amplifies shear-induced changes of the polymer conformations with enhanced stretching along the flow direction and a respective shrinkage in the transverse direction, as well as its rheological properties with a very strong shear thinning and a slow-downed tumbling. {\cblue Specifically, the power laws for the tumbling frequency and the shear-thinning behavior change qualitatively. In the regime, where activity dominates over shear, the tumbling frequency increases by a power-law with the exponent of 1/3 with increasing shear rate--half the value of a passive polymer--, which corresponds to slow-down of tumbling. Simultaneously, the shear viscosity decreases with an exponent of approximately $4/3$ with increasing shear rate, significantly faster than for passive polymers.}
These properties approach activity-independent universal asymptotic limits for larger activities. An analytical theory supports our findings and shows excellent qualitative agreement with the simulation results. Hence, the shear response differs qualitative from that of passive polymers~\cite{huang2010semidilute,pincus2023dilute}, of active polymers composed to active Brownian monomers~\cite{kaiser2014unusual,winkler2020physics,eisenstecken2017internal,anand2020conformation,martin2018active,panda2023characteristic}, and of active polar ring polymers~\cite{winkler2024active,kumar2024ringshear}. This reveals a particular coupling between polar propulsion and shear flow of such linear polymers, as is also reflected in the shear stress of entangled solutions~\cite{liverpool2001viscoelasticity,breoni2023giantactivity}.  

\section{Simulation Model}

The linear polymer comprises $N_m$ monomers at positions $\bm r_i$ ($i \in \{1,\ldots, N_m \}$), where consecutive monomers are connected by the harmonic spring potential
\begin{equation} \label{eq:bond_potential}
{U_s} = \frac{\kappa_s}{2} \sum_{i=1}^{N_m-1} (|\bm{r}_{i+1} - \bm{r}_i| - l_0)^2
\end{equation}
of finite equilibrium bond length $l_0$ and  $\kappa_s$ the force constant. Self-avoidance is taken into account by the repulsive, truncated, and shifted Lennard-Jones (LJ) potential 
\begin{align} \label{eq:ev_potential}
\displaystyle
U_{LJ} = \left\{ 
\begin{array}{cc} \displaystyle
  4 \epsilon\sum_{i>j}^{N_m} \left[\left(\frac{\sigma}{r_{ij}}\right)^{12} -\left(\frac{\sigma}{r_{ij}}\right)^{6} + \frac{1}{4}\right], & r_{ij} \leq r_c  \\ \displaystyle
 0 , & r_{ij} > r_c 
\end{array}
\right. ,
\end{align}
where $r_c = \sqrt[6]{2} \sigma$ is the cut-off distance between the monomers. Here, $\sigma$ is the LJ diameter of the monomers, $\epsilon$ the interaction strength, and ${r}_{ij} = |\bm{r}_j-\bm{r}_i|$ is the distance between monomer $i$ and $j$~\cite{anand2018structure,das2021coil,das2020introduction}. Hydrodynamic interactions are neglected, assuming a dry polymer environment.
The dynamics of the monomers is described by the overdamped Langevin equation
\begin{equation}
	 \bm{\dot{r}}_i(t) = \frac{1}{\zeta}\left[ \bm{F}_i(t)  + \bm {F}_i^a(t)    +\bm{\varGamma}_i(t)\right] + \mathbf{ K} \bm{r}_i(t),
	\label{Eq:eq_motion}
\end{equation}
where $\zeta$ is the friction coefficient,  $\bm{F_i}(t)$ denotes the conservative forces on monomer $i$ (Eqs.~\eqref{eq:bond_potential}, \eqref{eq:ev_potential}). Thermal noise is captured by the stochastic force $\bm{\varGamma}_i$,  a Gaussian white noise process with the first moment $\langle \bm{\varGamma}_i(t) \rangle = 0$  and the correlations  $\langle \bm{\varGamma}_i(t) \cdot \bm{\varGamma}_j(t') \rangle = 6 \zeta k_B T  \delta_{ij} \delta(t-t')$.
The second term in Eq.~\eqref{Eq:eq_motion}, ${\bm F}_i^a=f_a( \hat{\bm t}_{i+1}+ \hat{\bm t}_{i})/2$, is the active force of magnitude $f_a$. Here, $\hat{\bm t}_i = ({\bm r_{i}}- \bm r_{i-1})/|{\bm r_{i}}- \bm r_{i-1}|$ is the unit vector between monomers $i - 1$ and $i$. The active force on the first and last monomer is  $ f_a \, \hat{\bm t}_2/2$ and $f_a \, \hat {\bm t}_{N_m}/2$, respectively~\cite{anand2018structure}.

The last term in Eq.~\eqref{Eq:eq_motion} introduces a uniform linear shear flow, with the shear rate $\dot \gamma$, along the $x$-axis, and the gradient along the $y$-axis of the Cartesian reference frame, therefore, $K_{xy}=\dot \gamma$ is the only nonzero component of the shear rate tensor.

{\it Parameters:}
All parameters are presented in units of the LJ diameter $\sigma$, thermal energy $k_B T$, time $\tau = \zeta \sigma^2/(k_BT) = \sigma^2/D_t$, where $D_t=k_BT/\zeta$ is the translational diffusion coefficient of a free monomer, and force $k_B T/\sigma$. The spring constant $\kappa_s$ is chosen between $(5 \times 10^4- 10^5)k_B T/\sigma^2$ to avoid bond stretching under extreme limits of the active and shear forces. The bond length is $l_0=\sigma$, and the LJ energy as $\epsilon/k_B T =1$.  The dimensionless P\'eclet number $Pe= f_a l_0/ (k_B T)$, characterizing the strength of the active on a monomer, is varied from $0-100$. The time step $\Delta t$ is selected between $(10^{-7} - 10^{-4})\tau$. The flow strength is described by the dimensionless Weissenberg number $Wi_{Pe}=\dot{\gamma} \tau_r(Pe)$, where $\tau_r(Pe)$ as the longest polymer relaxation time in the absence of shear obtained from the end-to-end vector correlation function. Figure~\ref{Fig:relax} in App.~\ref{app:relaxation_time} provides an example of the relaxation time as a function of the P\'eclet number, showing a $1/Pe$ decay for $Pe \gtrsim 1$. Polymers with $N_m=200$, $400$, and $1000$ monomers are considered, corresponding to the contour length of $L=199l_0$, $399l_0$ and $999l_0$.

\section{Polymer conformations}

\begin{figure}[t]
    \centering
    \includegraphics[width=\columnwidth]{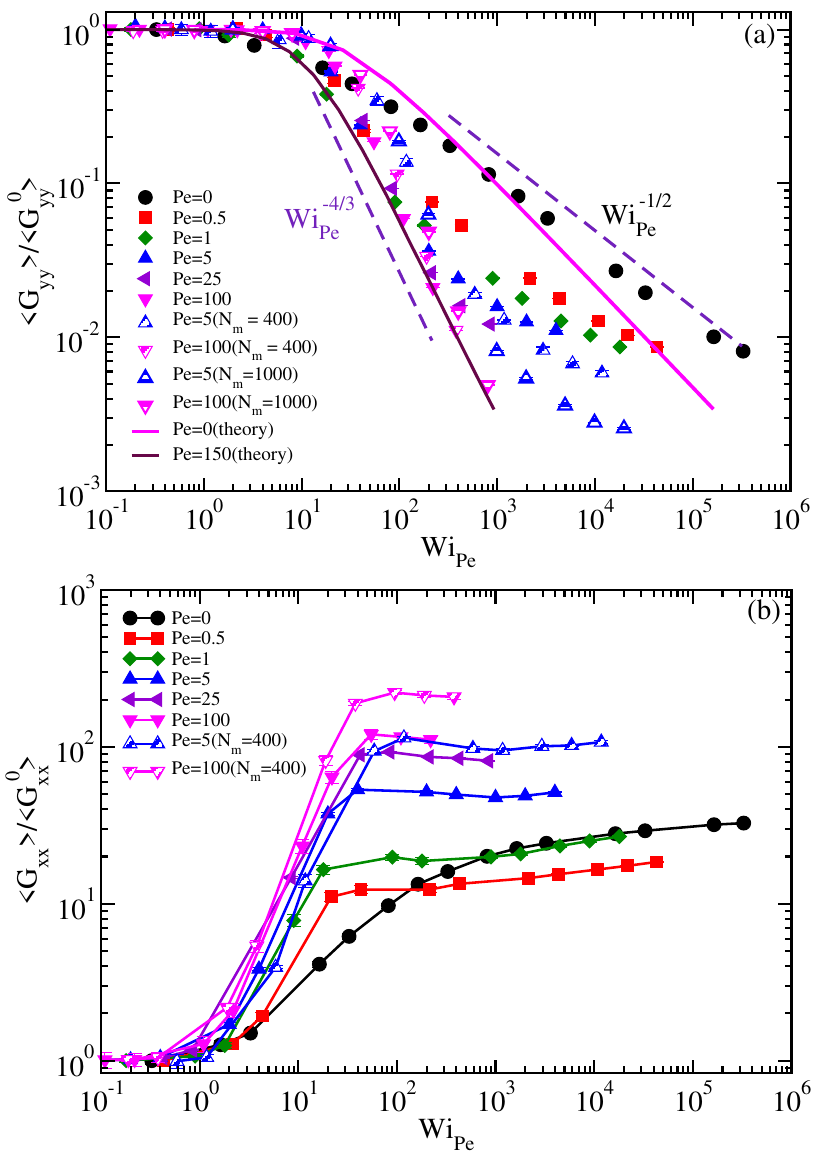}
    \caption{(a) Average radius-of-gyration tensor component $\langle G_{yy}\rangle $ in the gradient direction and (b) $\langle G_{xx}\rangle$ along the flow direction as a function of the Weissenberg number for various activities $Pe$ (legend) at $N_m=200$ (filled symbols), $N_m=400$, and $N_m = 1000$ (open symbols). In (a), the solid lines depict the theoretical shear-rate dependence obtained from Eq.~\eqref{eq:gyrat_theory} of App.~\ref{app:ananlytic} for passive polymers ($Pe = 0$, magenta) and active polymers ($Pe=150$, maroon).
    \cblue{ Here, $G_{yy}^0$ and $G_{xx}^0$ are the average radius-of-gyration components in absence of shear ($Wi_{Pe}=0$).}}
    \label{fig:gyy}
\end{figure}

The polymer conformations are characterized by the average, $\langle G_{\alpha \beta} \rangle$, of the instantaneous radius-of-gyration tensor
\begin{equation}
 G_{\alpha \beta} =  \frac{1}{N_m} \sum_{i=1}^{N_m}  \Delta{{r}_{i,\alpha}} \Delta{r_{i,\beta}} ,   
\end{equation}
where $\alpha, \beta \in \{x, y, z\}$ and $\Delta{\bm r_i}$ represents the position of the $i^{th}$ monomer relative to the polymer's center of mass.  Depending on the applied bond potential or implementation of the propulsion force, simulations reveal that polar active forces, in combination with excluded-volume interactions (EV), lead to a moderate~\cite{anand2018structure,fazelzadeh2023effects} or even strong~\cite{bianco2018globulelike,tejedor2024progressive} polymer collapse at large P\'eclet numbers ($Pe \gg  1$).  In contrast, a linear shear flow causes an average polymer stretching along the flow direction and simultaneous shrinkage in the transverse directions of passive polymers~\cite{teixeira2005shear,saha2012tumbling,winkler2010conformational,huang2010semidilute}.  
Fig.~\ref{fig:gyy}(a) shows the $y$-component of the radius-of-gyration tensor, $\langle G_{yy}\rangle/\langle G_{yy}^0\rangle$, where $\langle G_{yy}^0\rangle$ is the component in the absence of shear. It decreases with increasing Weissenberg number for $Wi_{Pe}>1$ and exhibits a pronounced dependence on the active force. 
For the passive polymer, $\langle G_{yy} \rangle$ follows the power law $\langle G_{yy}\rangle/\langle G_{yy}^0\rangle \sim Wi_{Pe}^{-1/2}$ in the large shear-rate limit, $Wi_{Pe} \gg 1$, in agreement with previous studies on passive polymers~\cite{schroeder2005characteristic,doyle1997dynamic,schroeder2018single,hur2001dynamics,shaqfeh2005dynamics,jendrejack2002stochastic,hsieh2004modeling,lyulin1999brownian,aust1999structure,colby2007shear,pincus2023dilute,tu2020direct}. The presence of activity substantially enhances the extent of shrinkage with increasing shear rate. Moreover, the curves for the various activities converge to a universal behavior irrespective of the P\'eclet number in the range $10 < Wi_{Pe} < 10^3$. Notably, the power-law exponent differs significantly from that of the passive case in this regime, and $\langle G_{yy}\rangle $ obeys the shear-rate dependence  $\langle G_{yy}\rangle/\langle G_{yy}^0\rangle \sim Wi_{Pe}^{-4/3}$. At high Weissenberg numbers, shear determines the nonequilibrium polymer conformations, and we observe a crossover from an activity-dominated to a shear-dominated regime, where the polymer properties coincide with those of a passive polymer.      


Analytical results of a Gaussian flexible active polar linear polymer model~\cite{philipps2022tangentially} (cf. App.~\ref{app:ananlytic}) agree qualitatively well with our simulation results. In particular, they confirm the strong dependence of the active polymer conformations on the Weissenberg number. The difference in the scaling exponent for $Pe=0$ may be a consequence of the absence of excluded-volume interactions in the analytical approach~\cite{winkler2010conformational,saha2012tumbling}. 

The shear-induced stretching of the radius-of-gyration tensor component $\langle G_{xx}\rangle/ \langle G_{xx}^0\rangle $ along the flow direction is illustrated in Fig.~\ref{fig:gyy}(b), where $\langle G_{xx}^0\rangle$ is the component in the absence of shear.  A passive polymer ($Pe = 0$)  exhibits a monotonically increasing stretching with increasing Weissenberg number. At very high shear rates, the polymer eventually stretches to its maximum extent~\cite{winkler2010conformational}. In the presence of activity, stretching increases faster with increasing shear rate compared to the passive case, and a plateau-like regime appears. A weak overshoot is observed for the active polymer before the plateau value is assumed. This may be due to the polymer's fast shrinkage in the transverse directions before shear dominates over activity and the polymer properties crossover to those of a passive polymer in shear flow.

\begin{figure}[t]
	\includegraphics[width=\columnwidth]{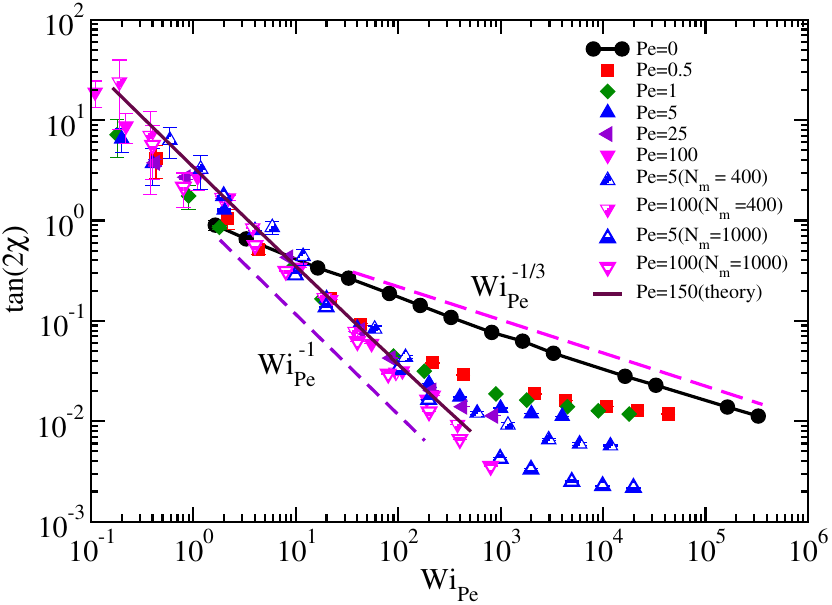}
	\caption{Alignment of the active polymer as a function of the Weissenberg number for various P\'eclet numbers. The solid maroon line is obtained from analytical theory for $Pe = 150$ and the polymer length $N_m = 200$. {\cblue The black solid line ($Wi_{Pe}>1$) is a guide for the eye of the data for a passive polymer.} The dashed lines indicate power laws.}
\label{Fig:alignment}
\end{figure}

\section{Alignment}

The stretching along the flow and shrinkage transverse to the flow direction lead to the preferred alignment of the average polymer shape. This alignment is characterized by the angle $\chi$, which is the angle between the eigenvector of the gyration tensor with the largest eigenvalue and the flow direction. It is obtained from the components of the radius of the gyration tensor via~\cite{aust1999structure,huang2010semidilute}
\begin{equation}
    \tan(2\chi) = \frac{2\langle G_{xy}\rangle}{\langle G_{xx}\rangle -\langle G_{yy}\rangle}.
\end{equation}
Figure~\ref{Fig:alignment} illustrates the alignment as a function of the shear rate. With the increasing shear flow, a passive polymer aligns along the flow direction, and $\tan(2 \chi)$ decreases with increasing shear rate with the power law $Wi_{Pe}^{-1/3}$ for $Wi_{Pe} \gg 1$\cite{huang2010semidilute,winkler2010conformational}. With increasing activity and flow strength, the alignment of the polymer enhances strongly, and $\tan(2 \chi)$ follows the scaling law  $Wi_{Pe}^{-1}$ in an intermediate shear-rate regime $10<Wi_{Pe}<10^3$ before it crosses over to the passive polymer behavior. The simulation results agree with theoretical predictions with nearly the same power-law exponent. 

\section{Inhomogeneous Stretching}
Polar active polymers have been shown to exhibit an inhomogeneous stretching of sub-strands along the polymer contour, thus, to a fore-aft-symmetry breaking~\cite{tejedor2024progressive}. The local stretching is quantified by the square root of the mean-square local distance $r_s = \sqrt{\langle \Delta \bm r_n^2(s) \rangle}= \sqrt{\langle ({\bm r}_{s+n}-{\bm r}_{s})^2 \rangle}$ (R-MSLD) between the monomers $s+n$ and $s$ ($n>0$, $s=1,\ldots,N_m-n$). 

In absence of activity ($Pe=0$), a polymer exhibits a self-similar structure along its contour, aside form end effects (Fig.~\ref{fig:inhom_stretch}(a) in App.~\ref{app:inhom_stretching}). The stretching of the segments remains symmetric with respect to the polymer ends, it increases with increasing shear rate, and asymptotically approaches a limiting value for very high Weissenberg numbers. In absence of shear, activity breaks the fore-aft symmetry, as displayed in Fig.~\ref{fig:inhom_stretch}(b) in App.~\ref{app:inhom_stretching}. Activity causes an overall compression of the polymer with increasing activity, where the head part is significantly stronger compressed than the tail region. This behavior is amplified with increasing P\'eclet number.

\begin{figure}[t]
    \centering
    \includegraphics[width=\columnwidth]{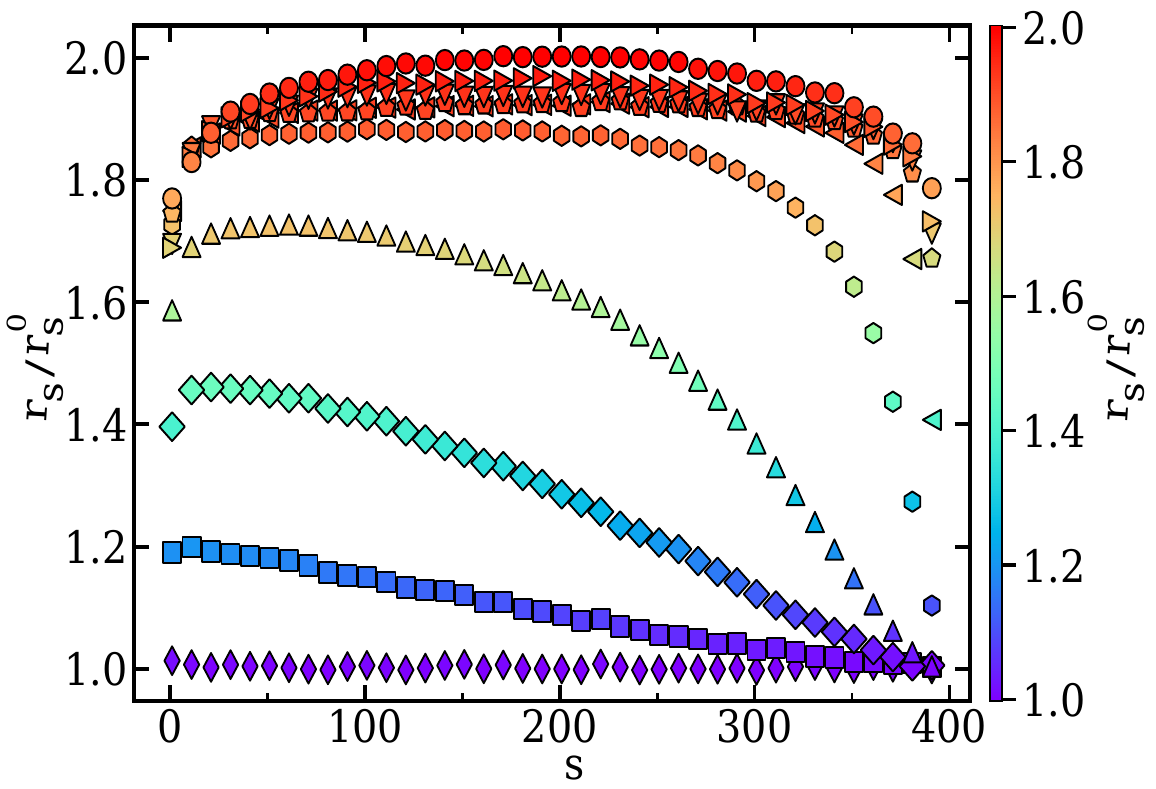}
    \caption{Square root of the mean-square local distance $r_s$ (R-MSLD) of polymer segments of length $n=9$ as a function of the monomer $s$ for $Pe=5$ and the Weissenberg numbers $Wi_{Pe} = 1.2$, $5.9$, $11.8$, $23.6$, $59$, $118$, $1180$, $2950$, $5900$, $11800$ (bottom to top). $r_s^0$ is the R-MSLD in absence of shear, and the colors indicate the relative magnitude of stretching.} 
    \label{fig:segment}
\end{figure}

Figure~\ref{fig:segment} illustrates the combined effect of shear flow and activity. In the activity-dominated regime ($Wi_{Pe} \lesssim 10^{2})$, shear enhances the anisotropic shrinkage, with a strong shrinkage toward the head end of the polymer. For small Weissenberg numbers $Wi_{Pe} \lesssim 20$, the R-MSLD decreases nearly linearly with increasing $s$ from the tail to the head. With increasing shear rate, a nearly symmetric R-MSLD develops with respect to the middle of the polymer. Here, shear dominates over activity, and the R-MSLD becomes similar to that of a passive polymer (cf. Fig.~\ref{fig:inhom_stretch}(a)).

The strong variation in the R-MSDL for small Weissenberg numbers indicates a breakdown of a self-similar polymer structure. Since the latter is the basis of many derived scaling laws in polymer physics~\cite{degennes1979scaling,doi1988theory}, the lack of self-similarity renders applications of such concepts difficult, and alternative approaches may have to be applied.

\begin{figure}[t]
    \centering
    \includegraphics[width=\columnwidth]{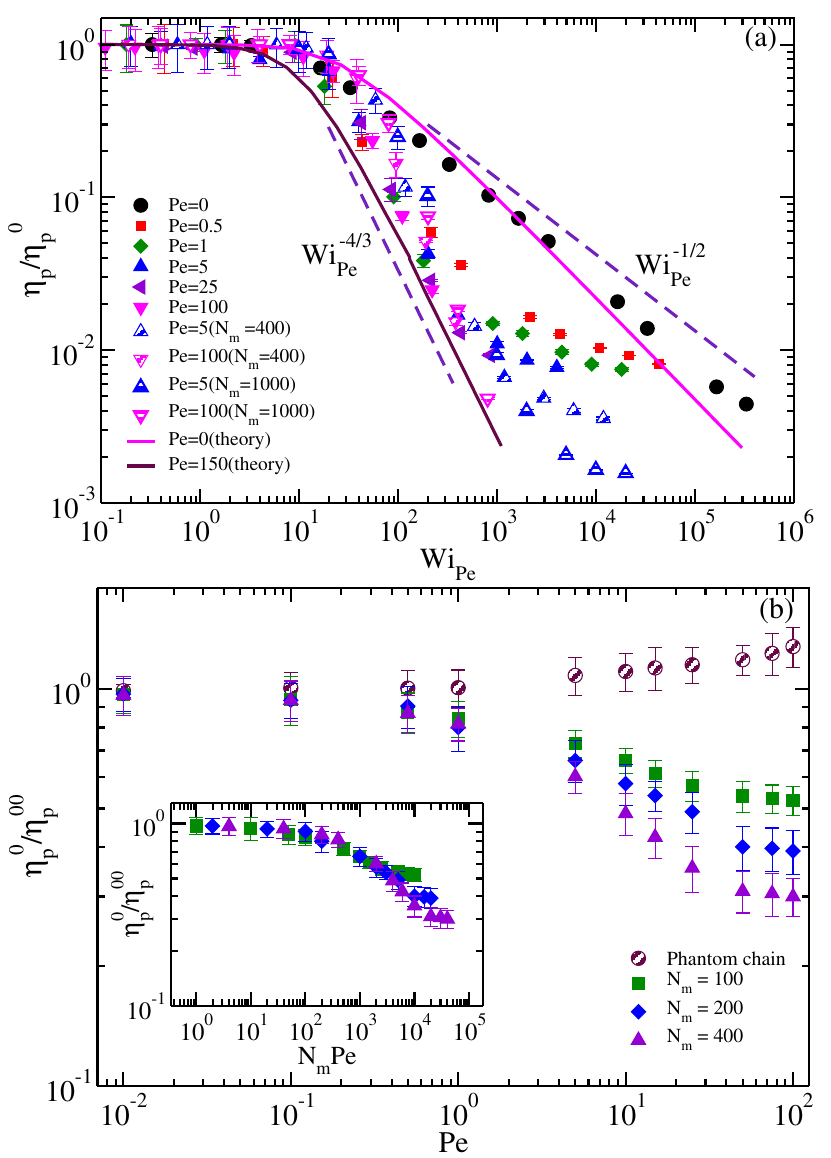}
    \caption{(a) Normalized shear viscosity $\eta_p/\eta_p^0$ of APLPs  as a function of $Wi_{Pe}$ for various activities $Pe$ (legend); $\eta_p^0$ is the $Pe$-dependent zero-shear viscosity.  The solid lines are derived from analytical theory (Eq.~\eqref{eq:eta}) for the P\'eclet numbers $Pe=0$ (magenta line) and  $Pe=150$ (maroon line).  
    (b) Normalized zero-shear viscosity  $\eta_p^0/\eta_p^{00}$ as a function of $Pe$ for active phantom polymers of length $N_m=200$ (top), and active polymers with EV interactions of length  $N_m= 100$, $200$, and $400$. $\eta_p^{00}$ is the zero-shear viscosity in the absence of activity. The inset displays a universal behavior of $\eta_p^0/\eta_p^{00}$ as a function of $N_m Pe$ for various polymer lengths.  
    }
    \label{fig:visc}
\end{figure}

\section{Shear Viscosity}

We characterize the rheological properties of the polymer by the intrinsic viscosity ($\eta_p$), which can be computed from the virial stress via  $\eta_p = |\sigma_{xy}| / \dot \gamma$, where the shear stress is $\sigma_{xy} = -\sum_{i=1}^{N_m} \langle (F_{xi}+F_{xi}^a) r_{yi} \rangle/V$~\cite{huang2010semidilute,winkler2010conformational}. 

Figure~\ref{fig:visc}(a) depicts the normalized shear viscosity $\eta_p/\eta_p^0$ as a function of the Weissenberg number, relative to the activity-dependent zero-shear viscosity $\eta_p ^0$. In the absence of activity, it exhibits the well established shear-thinning behavior with the power-law decrease $\eta_p /\eta_p^0 \sim Wi_{Pe}^{-1/2}$ as the flow strength increases ($Wi_{Pe} \gg 1$)  \cite{doyle1997dynamic,larson2005rheology,saha2012tumbling,huang2012non,panda2023characteristic}. The presence of the active forces significantly enhances the shear-thinning behavior, and the viscosity decreases much faster with increasing shear rate, with a gradual crossover from the passive to an asymptotic, $Pe$ independent limit with increasing activity. In the range $10<Wi_{Pe}<10^3$, the viscosity drops approximately with the power-law $\eta_p /\eta_p^0 \sim Wi_{Pe}^{-4/3}$, with a strikingly large exponent $4/3$. In the limit of $Wi_{Pe} \gg 1$, shear dominates over the activity, and the shear viscosity exhibits a crossover from the activity-dominant regime to the shear-dominant regime of a passive polymer.  {\cblue We like to emphasize that the average radius-of-gyration tensor component $\langle G_{yy}\rangle $ and the viscosity $\eta_p$ exhibit the same shear-rate dependence. The various scaling regimes appear nearly at the same Weissenberg number range for for both properties. Importantly, the ratio of these two normalized quantities,  \( \langle G_{yy} \rangle \) to \( \eta_p \),  varies between $25\%$ and $35\%$ as \( Wi_{Pe} \) changes over more than four orders of magnitude, underlining that both quantities exhibit a similar behavior.
} Additionally, the two quantities exhibit an identical shear-rate dependence even in the presence of Gaussian fluctuations in the active force with the same mean $Pe$ as illustrated in Fig.~\ref{fig:fluctuation} of App.~\ref{app:viscosity}.   

The analytical approach captures the activity-dependence of the shear viscosity very well. It also reflects the similarity in the shear-rate dependence of the radius-of-gyration component $\langle G_{yy}\rangle $ and the viscosity $\eta_p$.

Activity leads to a moderate collapse of an APLP in the presence of excluded-volume interactions at intermediate $Pe$ and a $Pe$-independent size for $Pe \gg 1$~\cite{anand2018structure,fazelzadeh2023effects}. This implies a reduction of the zero-shear viscosity $\eta_p^0$ as illustrated Fig.~\ref{fig:visc}(b), where the activity-dependence of $\eta_p^0$  is extracted from the plateau regime of the shear viscosity in the limit $Wi_{Pe} \ll 1$. The zero-shear viscosity in the presence of EV decreases with increasing activity and approaches a plateau for large activities $Pe\gg1 $. Notably, the polymer viscosity is reduced by about a factor of two. A qualitatively similar decrease of the zero-shear viscosity has been found for semiflexible active Brownian polymers (ABPO)~
\cite{martin2018active,anand2020conformation,panda2023characteristic}.
The qualitatively distinct activity-dependence of the zero-shear viscosity of the phantom polymer (absence of EV interactions) -- is nearly $Pe$ independent -- emphasizes the importance of excluded-volume interactions for polymers under shear flow. 

Interestingly, the zero-shear viscosity of the active polymer attains a length-independent universal behavior over a range of P\'eclet numbers as a function of $N_mPe$, as indicated in the inset of  Fig.~\ref{fig:visc}(b). A similar scaling of the structural properties in absence of shear is obtained in Ref.~\cite{tejedor2024progressive}, and in Refs.~\cite{philipps2022tangentially,bianco2018globulelike} for the diffusion coefficient.


\begin{figure}[t]
    \centering
     \includegraphics[width=0.9\columnwidth]{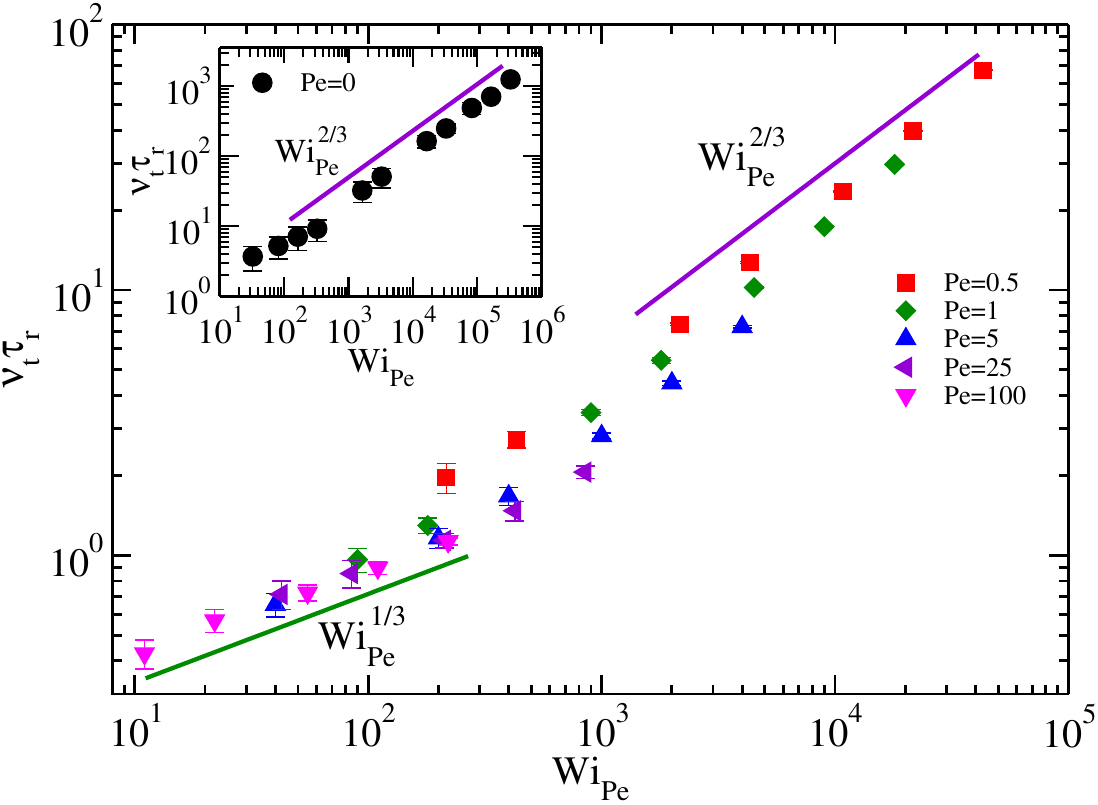}
    \caption{Normalized tumbling frequency $\nu_t \tau_r(Pe)$   of an APLP as a function of the Weissenberg number for various $Pe$ (legend) at $N_m=200$, where $\tau_r(Pe)$ is the end-to-end vector relaxation time. The solid green and purple lines indicate the power-law behavior with the exponents $1/3$ (activity-dominant) and 2/3  (shear-dominant). See supporting movies S1, S2, and S3~\cite{supp}. The inset displays the tumbling frequency for the passive polymer. }
    \label{fig:tumb}
\end{figure}

\section{Tumbling}

Passive linear polymers under shear flow exhibit a tumbling motion~\cite{schroeder2005characteristic,winkler2006semiflexible,huang2012non,singh2019tumbling,singh2020flow,li2021tumbling,winkler2024active}, for which analytical theory and simulations predict an associated characteristic time, the tumbling time $\tau_t$, with the shear-rate dependence $\tau_t \sim {\dot \gamma}^{-2/3}$~\cite{schroeder2005characteristic,winkler2006semiflexible,huang2012non,singh2019tumbling,singh2020flow,li2021tumbling,winkler2024active,saha2012tumbling,lang2014dynamics}. We determine the tumbling time in two ways: via the time-dependence of the projection of the polymer end-to-end vector onto the $x$-axis of the Cartesian reference frame~\cite{singh2020flow}
\begin{equation} \label{eq:cost}
    \cos(\theta) = \frac{\bm R_e \cdot \bm e_x }{|\bm R_e|} ,
\end{equation} 
and the cross-correlation function of radius-of-gyration tensor components~\cite{huang2012non,singh2020flow,schroeder2005characteristic,tu2020direct,panda2023characteristic}
\begin{equation} \label{eq:corr}
C_{xy}(t)=\frac{\lla \Delta G_{xx}(t_0) \Delta G_{yy}(t_0+t)\rra}{\sqrt{\lla \Delta G^2_{xx}(t_0) \rra \lla \Delta G^2_{yy}(t_0)\rra}},    
\end{equation}
where $\Delta G_{\beta \beta}(t) = G_{\beta \beta}(t) - \langle G_{\beta \beta}\rangle $ is the deviation in the gyration tensor from its average value $\langle G_{\beta \beta} \rangle$.
Both approaches yield the same shear-rate dependence. Figure~\ref{fig:tumbling} in App.~\ref{app:tumbling} provides examples of the projection Eq.~\eqref{eq:cost} and the correlation function Eq.~\eqref{eq:corr}. 

The projection in Fig.~\ref{fig:tumbling}(a) (App.~\ref{app:tumbling}) shows a cyclic motion of varying time intervals between successive crossings of the zero shear plane and allows for a convenient calculation of an average tumbling time $\tau_t$. For numerical convenience, we choose the value $\cos(\theta) = 0.85$ to define the tumbling interval.

The tumbling times extracted from both approaches agree with each other. Since the times obtained from the projection show less scatter, we discuss and present those only. The inverse of the average tumbling time, $\tau_t$, -- the tumbling frequency  $\nu_t =1/\tau_t$ -- is displayed in Fig.~\ref{fig:tumb} as a function of the Weissenberg number and for various $Pe$. Noteworthy, for the active polymer ($Pe>1$), the tumbling frequency exhibits the power-law increase $Wi_{Pe}^{1/3}$ with increasing shear rate in the regime $10<Wi_{Pe}<10^3$, which is substantially slower than that of a passive polymer. At larger Weissenberg numbers, a crossover to the passive limit occurs (cf. movies S1, S2, and S3~\cite{supp}). The latter is consistent with the crossover in radius-of-gyration tensor $\langle G_{yy} \rangle$ (Fig.~\ref{fig:gyy}). We attribute the smaller exponent in the intermediate shear-rate regime to the stronger shrinkage of the polymer in the gradient direction. The smaller object experiences a weaker shear flow, so transport along the shear direction requires longer. This is consistent with the observed enhanced shear thinning. This argument is supported by the following scaling argument~\cite{saha2012tumbling}. During a tumbling event, a monomer at the position $y_t \approx \sqrt{\langle G_{yy}\rangle}$ is convected by the flow over distance $L_t< N l$ with the velocity $v_x \approx \dot \gamma y_t$. Hence, $\tau_t \sim L_t/v_x \sim (\dot \gamma \sqrt{\langle G_{yy}\rangle})^{-1} \sim (\dot \gamma)^{-1/3}$, in agreement with the simulation results and the analytical model.

\section{Summary}

Using coarse-grained computer simulations and analytical theory, we have presented a comprehensive study of the conformational and rheological properties of self-avoiding active polar linear polymers under linear shear flow. Our studies reveal a strong impact of  polar activity on the behavior of flexible linear polymers in shear flow. A significantly enhanced shear-induced stretching along the flow direction and a respective shrinkage in the transverse direction for intermediate shear rates implies strong shear thinning.  The shear viscosity of polymer decreases as $\eta_p \sim Wi_{Pe}^{-4/3}$ with an exponent $(4/3)$ much larger than that of a passive polymer in dilute solution~\cite{huang2010semidilute,saha2012tumbling}. This shear thinning is much more pronounced compared to that of flexible active Brownian polymers, where $\eta_p \sim Wi_{Pe}^{-3/4}$ is predicted~\cite{martin2018active,panda2023characteristic}, which is related to the enhanced persistent motion along the polymer contour of the shear-aligned APLP. Analytical theory supports our findings, showing excellent qualitative agreement with the simulation results for various physical quantities, such as shear-induced compression,  alignment of the polymer, and shear viscosity.

 Furthermore, our simulations demonstrated that the active polar polymer's contribution to viscosity decreases significantly, becoming nearly an order of magnitude smaller than that of a passive polymer at the same Weissenberg number. This decrease in the viscosity of the polymeric solution is purely driven by the polar nature of the activity and its coupling with the conformations and excluded-volume interactions. This is supported by simulations of a phantom polymer, which indicates that excluded volume is the primary reason behind the decrease in the zero-shear viscosity, as the viscosity of the phantom active polymer merely changes with activity~\cite{panda2023characteristic}. Additionally, the tumbling frequency of the APLP shows a smaller power-law exponent than the passive polymer and grows as $Wi^{1/3}$ with increasing Weissenberg number as compared to $Wi^{2/3}$ of a passive polymer. The smaller exponent is related to the strong shrinkage of the polymer in the gradient direction because smaller objects experience a weaker shear flow.

 Experiments on {\em Tubifex} worms exhibit a qualitatively different trend in the activity-dependent shear viscosity \cite{deblais2020rheology} with increasing activity, namely a reduced shear-thinning behavior.  Here, entanglements may play a significant role, whereas our study has addressed the polymer properties in a dilute solution.  
Further theoretical endeavors are necessary to resolve the influence of collective effects on the rheology of APLPs and to unravel specificities in systems such as 
{\it Lumbriculus variegatus}~\cite{deblais2023worm,nguyen2021emergent,ozkan2021collective},  {\it C. elegans}~\cite{ding2019shared,backholm2013viscoelastic}, and {\it Turbatrix aceti nematodes}~\cite{ali2023oscillating}.

Computer simulations of passive polymers reveal that hydrodynamic interactions can quantitatively influence the mechanical and transport properties, but the qualitative behavior is very similar~\cite{huang2010semidilute}.  Analogously, simulations of semiflexible active polymers with and without hydrodynamic interactions lead to the same conclusion~\cite{saju2024dry}. In the latter simulations of dry active polymers, the active force is applied normal to the polymer backbone rather than along the backbone as in our simulations. Hence, for a quantitative comparison with possible experiments, hydrodynamics simulations are desirable.      

Independent of that, our results suggest a root to the development of functional materials with activity-controlled rheological properties.

\begin{acknowledgments}
SPS  and RGW acknowledge the helpful discussion with M. Ripoll. SPS acknowledges funding support from the DST-SERB Grant No. CRG/2020/000661, and computational time at IISER Bhopal and Param Himalaya NSM facility.
\end{acknowledgments}

\appendix

\section{Activity dependence of relaxation time} \label{app:relaxation_time}

\begin{figure}[ht]
	\includegraphics[width=\columnwidth]{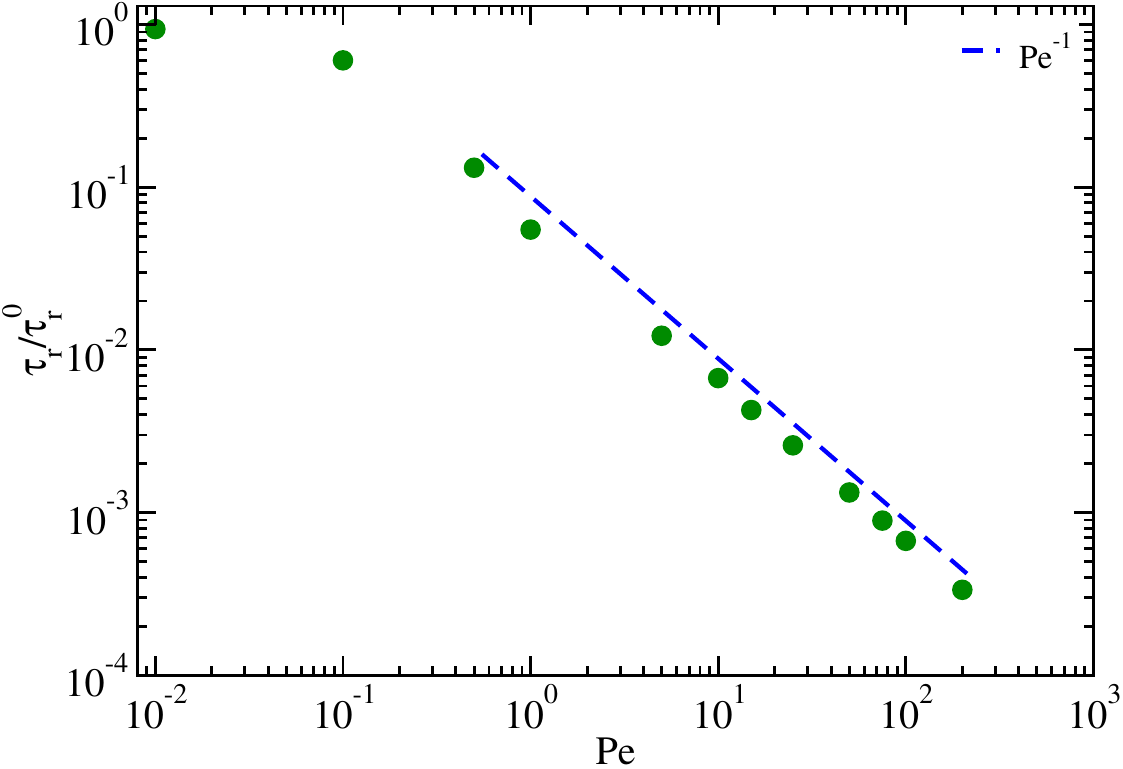}
	\caption{End-to-end vector relaxation time of the active polymer as a function of the P\'eclet number for the polymer length $N_m = 200$ at $Wi_{Pe} =0$. The dashed blue line indicates the power-law $Pe^{-1}$, and $\tau_r^0$ is the relaxation time in the absence of activity. }
\label{Fig:relax}
\end{figure}

The relaxation time of the polymer is determined from the autocorrelation function of the end-to-end vector, \( C_e(t) = \langle {\bm R}_e(t) \cdot {\bm R}_e(0) \rangle \), where ${\bm R}_e$ is the end-to-end vector.  The estimated correlation function of the APLP decays in a non-exponential manner~\cite{tejedor2024progressive}; therefore,  we define $\tau_r$ as the time where the correlation function has decayed to $1/e$, i.e., $C(\tau_r)=1/e$. 
Figure~\ref{Fig:relax} presents the active polymer's relaxation time $\tau_r$ as a function of the P\'eclet number $Pe$. For $Pe>1$, the relaxation time decays as $1/Pe$~\cite{fazelzadeh2023effects}.    



\section{Inhomogeneous stretching} \label{app:inhom_stretching}

\begin{figure}[t]
	\includegraphics[width=\columnwidth]{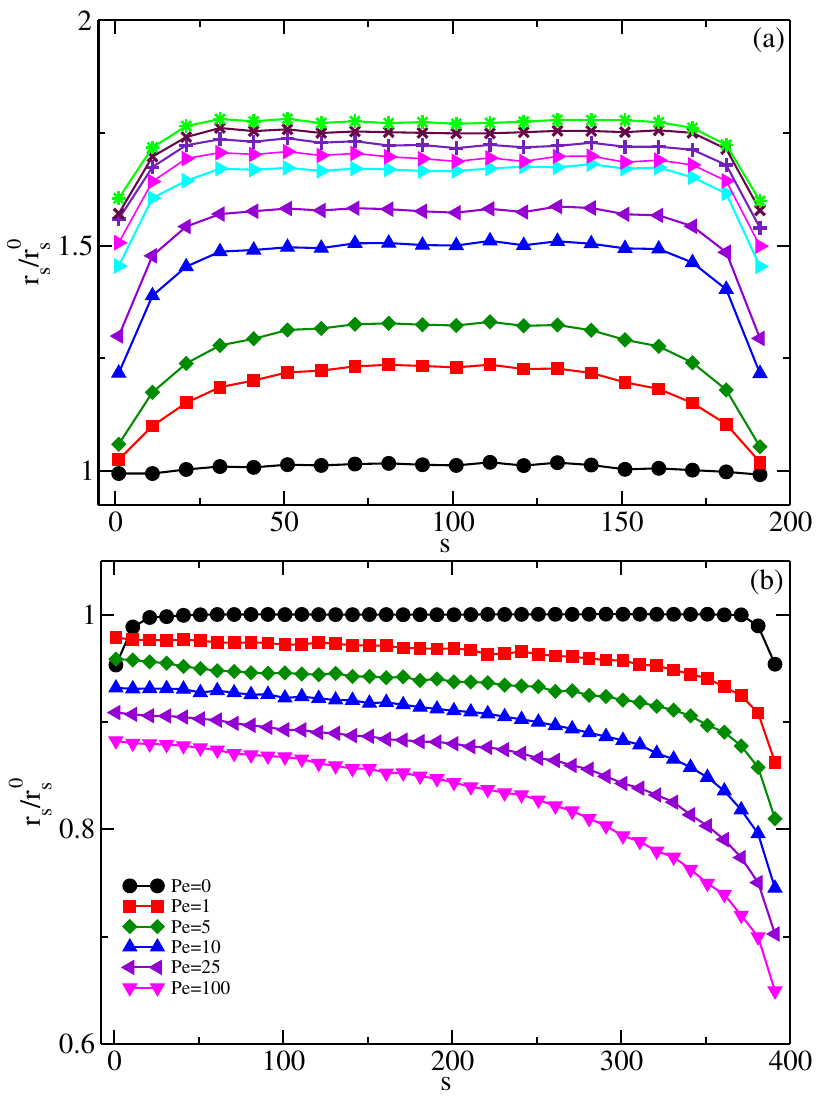}
	\caption{Root-mean-square local distance (R-MSLD) as a function of the monomer position $s$. (a) Passive polymer under shear flow for the Weissenberg numbers $Wi_{0}=32.8$, $164$, $328$, $1640$, $3280$, $16400$, $32800$, $82000$, $164000$, and $328000$ (bottom to top) for the number of monomers $N_m=200$. (b) Active polymer in absence of shear flow for various P\'eclet numbers (legend) and $N_m=400$.
    The $r_s^0$ is the R-MSLD in the absence of shear. 
    }
\label{fig:inhom_stretch}
\end{figure}

Inhomogeneous stretching (R-MSLD) of passive polymers under shear flow (Fig.~\ref{fig:inhom_stretch}(a)) and active polymers in absence of shear flow (Fig.~\ref{fig:inhom_stretch}(b).

\section{Viscosity -- Fluctuating activity} \label{app:viscosity}

\begin{figure}[t]
	\includegraphics[width=\columnwidth]{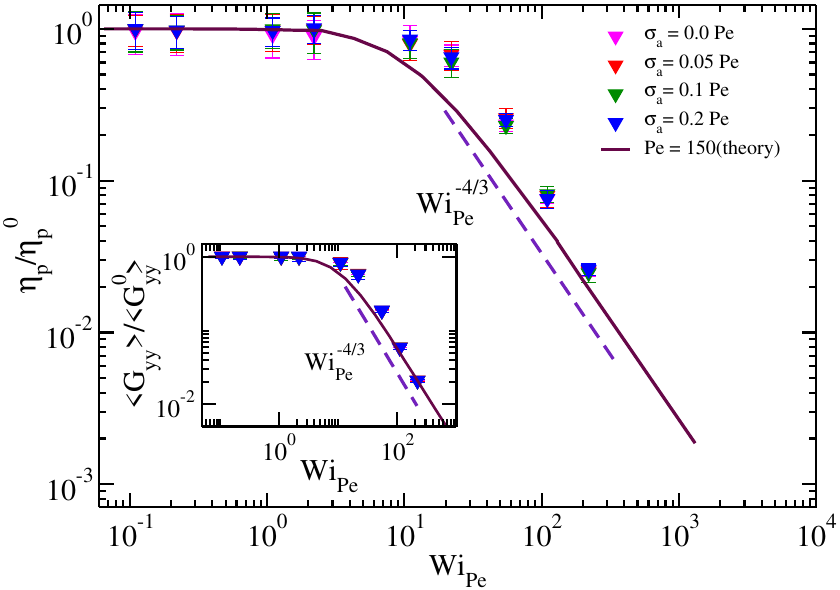}
	\caption{Shear viscosity, $\eta_p/\eta_p^0$, of APLPs as a function of the Weissenberg number $Wi_{Pe}$ for various activity-force fluctuations {\cblue $\sigma_{a} = 0.0\,Pe$, $0.05\,Pe$, $0.1\,Pe$, and $0.2\,Pe$ at $Pe = 100$}. The inset shows the normalized gradient-direction component of the radius-of-gyration tensor as a function of $Wi_{Pe}$ for  $Pe = 100$ and the same fluctuations. }
\label{fig:fluctuation}
\end{figure}

The influence of possible fluctuations in the active forces along the bonds is presented in Fig.~\ref{fig:fluctuation} for the viscosity and the radius of gyration along the gradient direction. The fluctuations are captured by adding a Gaussian noise term to the constant force $f_a$, i.e., we consider the force $f_{a_i} + \delta f_{a_i}$ on the $i$th monomer.  The fluctuating force  $\delta f_{a_i}$ is of zero mean, and the second moment obeys the relation $\langle \delta f_{a_i}(t) \delta f_{a_j}(t') \rangle =2\sigma_{a}^2 \delta_{ij}\delta(t-t')$. As shown in Fig.~\ref{fig:fluctuation} for the values $\sigma_{a} = 0.0\,Pe$, $0.05\,Pe$, $0.1\,Pe$, and $0.2\,Pe$ {\cblue at $Pe=100$}, no significant changes in the viscosity or gyration tensor component are obtained.  

\section{Tumbling}\label{app:tumbling}

\begin{figure}[t]
    \centering
     \includegraphics[width=\columnwidth]{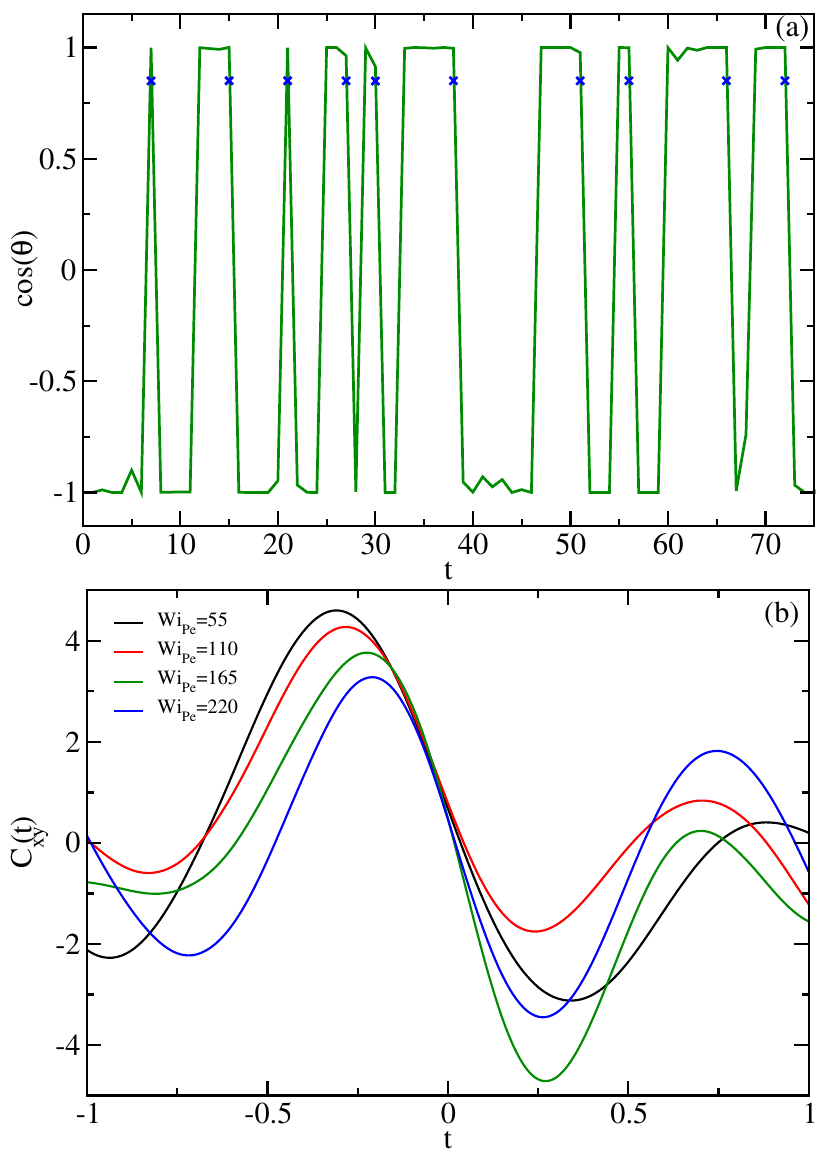}
     
    \caption{(a) Time dependence of the projection of the polymer end-to-end vector onto the
    $x$-axis of the Cartesian reference frame, $\cos(\theta)$, (Eq.~\eqref{eq:cost}) for $Pe = 100$ and $Wi_{Pe}$ = 220. The crosses indicate the times where $\cos(\theta) =0.85$, which are used in the calculation of a tumbling time.
   (b) Cross-correlation function $C_{xy}(t)$ (Eq.~\eqref{eq:corr}) as a function of time  for $N_m = 200$, $Pe = 100$, and various Weissenberg numbers (legend).  }
    \label{fig:tumbling}
\end{figure}

The projection of the unit polymer end-to-end vector onto the flow direction ($\cos(\theta)$) exhibits as a cyclic motion with time $t$ as displayed in Fig.~\ref{fig:tumbling}(a). The successive crossings of the zero shear plane allows for a convenient calculation of an average tumbling time $\tau_t$.
 
Figure~\ref{fig:tumbling}(b) illustrates the oscillatory behavior of the cross-correlation function $C_{xy}(t)$ (Eq.~\eqref{eq:corr}). Each curve exhibits a deep minimum at time $t_+ >0$ and a pronounced peak at time $t_-<0$ and decays to zero at larger time-lags. The difference between the maximum ($t_-$) and minimum ($t_+$) defines the average time required for the polymer to complete a cycle of conformational change. The minimum ($t_+$) arises from the correlations between stretching along the flow and compression in the gradient direction occurring later. Similarly, a maximum ($t_-$) resulting from stretching in the flow direction with swelling earlier in the gradient direction or vice versa \cite{huang2010semidilute,huang2012non}. Hence, the tumbling event is characterized by the time $\tau_t = 2(t_+ - t_-)$. Factor two is introduced because two non-equivalent conformations lead to a maximum and a minimum, respectively, and will be (more or less) assumed during a cycle.

\section{Analytical model for  APLP under shear flow} \label{app:ananlytic}

A flexible discrete linear polymer is composed of beads connected by harmonic bonds \cite{philipps2022tangentially}. The beads at positions $\bm r_i (t)$, $i \in\{0,\ldots N\}$, evolve in time ($t$) according to the Langevin equations \cite{philipps2022tangentially}
\begin{align} \label{Eq:dis_eq1}
    \zeta \frac{d {\bm r}_{0}}{dt} = & \ \left(\frac{f_a}{2} + \frac{3k_BT\mu}{l^2}\right){\bm R}_1(t)+ \zeta \mathbf{K} {\bm r}_0 + {\bm \varGamma}_0  ,\\
\label{Eq:dis_eq2}
    \zeta \frac{d {\bm r}_{i}}{dt} = & \ \frac{f_a}{2} ( {\bm R}_{i+1}(t) + {\bm R}_{i}(t)) + \zeta \mathbf{K}  {\bm r}_i  \\ &  \nonumber + \frac{3k_BT\mu}{l^2} ( {\bm R}_{i+1}(t) - {\bm R}_{i}(t))
   + {\bm \varGamma}_j , \\ \label{Eq:dis_eq3}
    \zeta \frac{d {\bm r}_{N}}{dt} = & \ \left(\frac{f_a}{2} - \frac{3k_BT
    \mu}{l^2}\right) \bm R_N(t)+ \zeta \mathbf{K} {\bm r}_N + \bm \varGamma_N .
\end{align}
Here, $3k_BT \mu ({\bm R}_{i+1}-{\bm R}_i)/{l^2}$ is the bond force on the $i^{th}$ bead, with $\bm R_{i+1} = \bm r_{i+1} - \bm r_i$ the bond vector. The Lagrange multiplier $\mu$  enforces the global inextensibility constraint  
\begin{equation} \label{eq:constraint}
    N l^2=  \sum_{i=1}^{N} \lla \bm R_i^2(t) \rra.
\end{equation}

The equations of motion are solved by an eigenfunction expansion in terms of a biorthonormal basis set ${\bm b}_n=({\bm b}_n^0,{\bm b}_n^1,\ldots,{\bm b}_n^N)^T$ and ${\bm b}_n^{\dag}=({\bm b}_n^{0^\dag},{\bm b}_n^{1\dag},\ldots,{\bm b}_n^{N\dag})^T$ of a nonsymmetric matrix $\mathbf{M}$ and its adjunct, corresponding to Eq.~\eqref{Eq:dis_eq1} \cite{philipps2022tangentially}, i.e.,
$r_{\alpha \, i}=\sum_{n=0}^{N}\chi_{\alpha n}(t) b_n^i$  and $\varGamma_i=\sum_{n=0}^{N}\Tilde\varGamma_{\alpha n}(t)b_n^i$, with the mode amplitudes $\chi_{\alpha n}$ and $\Tilde\varGamma_{\alpha n}$, respectively. 
Explicitly, the basis functions are given by
\begin{align} 
	&b_n^{i} = \sqrt{\frac{2}{N+1}} \frac{e^{-d i}}{\sqrt{(1-r_d^2) \sin^2 k_n + (r_d - \lambda_n/2)^2}} 
	\label{align:eigen_dis_b}
	\\		
	& \times \Big[ \sqrt{1-r_d^2} \sin k_n \cos(k_n i) + (r_d - \lambda_n/2) \sin(k_n i) \Big], \nonumber
	\\[0.5em]
	&b_n^{i^{\dagger}} = e^{2 d i} b_n^{i}, \hspace{1cm} m \in \mathbb{N}_0,
	\label{align:eigen_dis_b_dag}
	\\[0.5em] 
	&b_0^{i} = \sqrt{\frac{\sinh(d)}{e^{dN}\sinh(d(N+1))}}, 
	\label{align:eigen_dis_norm}
	\\[0.5em]
	&d = \ln (\sqrt{1+r_d}/\sqrt{1-r_d}), 
	\hspace{1.44cm}
	r_d <1 , \\
	&d = \ln (\sqrt{1+r_d}/\sqrt{r_d-1}) - i \frac{\pi}{2}, 
	\hspace{0.6cm}
	r_d >1 , 
\end{align}	
with the wave numbers $k_n = n \pi / (N+1)$ and the abbreviation 
\begin{equation} \label{eq:abb_d}
    r_d = \frac{f_a l^2}{6 k_B T \mu} = \frac{Pe}{6 \mu} .    
\end{equation}
Insertion in Eq.~\eqref{Eq:dis_eq1}$-$\eqref{Eq:dis_eq3} yields the equations of motion for the mode amplitudes
\begin{equation} \label{eq:mode_eom}
    \zeta \frac{d \chi_{\alpha n}}{dt} = -\xi_{n}\chi_{\alpha n} + \Tilde\varGamma_{\alpha n}(t)+  \dot \gamma   \zeta\delta_{x \alpha }\chi_{\alpha n}  ,
\end{equation}
with the eigenvalues $\xi_n$ of the matrix $\mathbf{M}$. They are given by
\begin{equation}
   \xi_n = \frac{6 k_BT \mu}{l^2} \left( 1 - \sqrt{1-r_d^2} \cos k_n \right) .
\end{equation}
The linear equations~\eqref{eq:mode_eom} can easily be solved~\cite{philipps2022tangentially}. 

The average gyration tensor is defined as
\begin{equation}
\langle {G}_{\alpha \beta}\rangle = \frac{1}{N+1} \sum_{i=0}^{N} \lla ({r}_{\alpha \, i}(t) -r_{\alpha \, cm}(t)) (r_{\beta \, i}(t) -{r}_{\beta \, cm}(t)) \rra .
\label{Eq:gyration}
\end{equation}
Insertion of the expansion yields  
\begin{align} \nonumber \label{eq:gyrat_theory}
\langle G_{\alpha\beta} \rangle = \frac{1}{N+1} &  \left\{ \displaystyle \sum_{m,n=1}^{N} \right. \lla \chi_{\alpha m}(0) \chi_{\beta n}(0) \rra \\ & \left. \times \left[ ({\bm b}_m\cdot {\bm b}_n)  -   \displaystyle  \sum_{i,j=0}^{N}(b_m^i b_n^j) \right] \right\} .
\end{align}

The contribution of the polymer to the shear viscosity is calculated via the viral stress
\begin{equation} \label{eq:shear_stress}
\sigma_{xy} = -\frac{1}{V}\sum_{i=0}^{N} \langle F_{x \, i } \, { r}_{y \, i} \rangle,
\end{equation}
where $F_{x i}$ is the sum of the bond and active force on the $i^{th}$ monomer, and $V$ is the system's volume.  
In terms of the eigenfunctions, Eq.~\eqref{eq:shear_stress} reads
\begin{equation}
\sigma_{xy} = - \frac{3 k_BT \mu}{\zeta l V} \sum_{n=1}^{N}\sum_{m=0}^{N}  \lla \chi_{xn}(0) \chi_{ym}(0)\rra
A_{nm}, 
\end{equation}
with 
\begin{align} \nonumber 
A_{nm} = & \ (b_n^1-b_n^0)b_m^0 + (b_n^{N-1}-b_n^N)b_m^N \\ & \  + \sum_{i=1}^{N-1} (b_n^{i+1} + b_n^{i-1} - 2b_n^i)b_m^i .
\end{align}
The shear viscosity follows from the relation $\eta_p= |\sigma_{xy}|/{\dot \gamma}$ as
\begin{equation}\label{eq:eta}
\eta = \frac{6 (k_BT)^2 \mu}{l^2 V} \sum_{n=1}^{N}\sum_{m=0}^{N} \frac{A_{nm}}{(\xi_n + \xi_m)^2} {\bm b}_n^{\dag} \cdot {\bm b}_m^{\dag} . 
\end{equation}
The shear-rate dependence of the viscosity is determined by $\mu(\dot \gamma)$. 

Analytical calculations are performed mainly for the polymer length $N = 200$, $r_d = 0$, and $r_d = 25$, where $r_d = 0$ corresponds to the P\'eclet number $Pe = 0$ and $r_d = 25$ corresponds to $Pe = 150$.


%

\end{document}